# System 2 Thinking in OpenAI's o1-Preview Model: Near-Perfect Performance on a Mathematics Exam

**J. C. F. de Winter \*, D. Dodou, Y. B. Eisma**

Faculty of Mechanical Engineering, Delft University of Technology, The Netherlands

\* Corresponding author: j.c.f.dewinter@tudelft.nl

**Abstract**
The processes underlying human cognition are often divided into System 1, which involves fast, intuitive thinking, and System 2, which involves slow, deliberate reasoning. Previously, large language models were criticized for lacking the deeper, more analytical capabilities of System 2. In September 2024, OpenAI introduced the *o1* model series, designed to handle System 2-like reasoning. While OpenAI's benchmarks are promising, independent validation is still needed. In this study, we tested the *o1-preview* model twice on the Dutch 'Mathematics B' final exam. It scored a near-perfect 76 and 74 out of 76 points. For context, only 24 out of 16,414 students in the Netherlands achieved a perfect score. By comparison, the *GPT-4o* model scored 66 and 62 out of 76, well above the Dutch students' average of 40.63 points. Neither model had access to the exam figures. Since there was a risk of model contamination (i.e., the knowledge cutoff for *o1-preview* and *GPT-4o* was after the exam was published online), we repeated the procedure with a new Mathematics B exam that was published after the cutoff date. The results again indicated that *o1-preview* performed strongly (97.8th percentile), which suggests that contamination was not a factor. We also show that there is some variability in the output of *o1-preview*, which means that sometimes there is 'luck' (the answer is correct) or 'bad luck' (the output has diverged into something that is incorrect). We demonstrate that the self-consistency approach, where repeated prompts are given and the most common answer is selected, is a useful strategy for identifying the correct answer. It is concluded that while OpenAI's new model series holds great potential, certain risks must be considered.

## 1. Introduction

Large language models (LLMs) became accessible to the public with the launch of ChatGPT in November 2022 [1]. This tool has impacted various fields, particularly education and academia. In education, ChatGPT has transformed teaching practices [2,3], while in academic research, many students and scholars now rely on the tool to assist with writing papers and reports [4–8]. Additionally, ChatGPT is widely used to support the peer review process for academic work [9,10].

At the beginning of 2023, OpenAI introduced *GPT-4* [11], a larger base model compared to the previous *GPT-3.5* model. *GPT-4*, along with further fine-tunings and iterative improvements (such as *GPT-4-turbo* and *GPT-4o*), has continued to lead the LLM rankings for a long period [12,13]. Other models, such as Anthropic's Claude [14], Google's Gemini [15], and Meta's Llama [16], are close behind and outperform *GPT-4* on some benchmarks [17,18].

A common criticism of LLMs, despite their already impressive performance, is that they seem to be reaching a plateau, with further improvements in output quality presumably requiring even larger base models or the use of more or higher-quality data [19,20]. Another frequent critique is that LLMs function as "stochastic parrots" [21], meaning they act as autoregressive token generators, where subsequent tokens are conditioned on the previously generated tokens. So far, LLMs have not often demonstrated self-reflection or the correction of their output, or a deeper understanding of the meaning behind the generated text. Indeed, while LLMs can produce fluent text, they struggle with seemingly basic tasks like counting [22], arithmetic [23,24], and reasoning [24,25]. A clear example is the Tower of Hanoi puzzle [26]. Although LLMs can explain how to solve such puzzles and can generate a computer program to do so, they perform poorly when asked to directly solve even simple versions of this puzzle [26].

The current limitations of LLMs can be better understood through the framework of System 1 and System 2 thinking, as outlined by Evans and Stanovich [27] and popularized by Kahneman [28] (Table 1). Traditional computer systems have historically excelled at System 2 processes, i.e., processes that require logic and algorithmic processing. Until the arrival of LLMs, computers struggled with System 1 processes, which are the intuitive and context-sensitive skills that humans excel at. The breakthrough of LLMs is their ability to handle System 1-like tasks, such as generating natural language fluently and responding to context. This achievement is noteworthy in light of Moravec's paradox [29,30], which says that tasks that humans find intuitive, such as language processing, have long been difficult for computers to perform. LLMs have overcome this limitation by excelling at tasks that mimic the automatic, context-driven nature of System 1 thinking.

Table 1. *Overview of human System 1 (automatic, fast, nonconscious) and System 2 (effortful, slow, conscious) cognitive processes in dual-process theory [38] (based on [27]).*

|  | **System 1 processes** | **System 2 processes** |  |
| --- | --- | --- | --- |
| 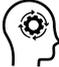 | Automatic | Effortful | 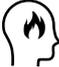 |
| 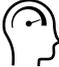 | Fast | Slow | 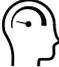 |
| 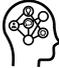 | Contextualized | Abstract | 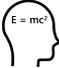 |
| 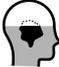 | Nonconscious | Conscious | 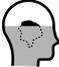 |
| 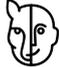 | Like animal cognition | Distinctively human | 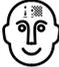 |
| 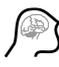 | Evolved early | Evolved late | 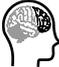 |

However, LLMs still lack the ability to perform System 2 reasoning, i.e., higher-order, deliberate thought processes. LLMs neither fully replicate human-like System 2 abilities nor possess the same reasoning capabilities as traditional computer systems. The next frontier is to improve LLMs with System 2 capabilities, which would allow them to combine the strengths of both systems: the intuitive natural language understanding of System 1 and the deep thinking of System 2. This combination would move LLMs closer to functioning like humans, who rely on the interplay of both systems to navigate complex tasks.

Li et al. [31] and Feng et al. [32] provided a theoretical foundation for the fact that chain-of-thought (CoT) reasoning is essential for solving mathematical and decision-making problems. Techniques to improve the outputs of LLMs through CoT or other forms of internal validation have been previously described by OpenAI (e.g., [33]) and Google/DeepMind [34,35]. Kamruzzaman and Kim [36] noted that traditional zero-shot prompting closely mirrors the intuitive processes of System 1, while CoT prompting resembles the more deliberate and analytical reasoning of System 2. As pointed out above, LLMs have struggled with abstract thinking, i.e., System 2 processes. However, this limitation of LLMs has diminished with OpenAI's *o1* model, released on 12 September 2024. Unlike previous LLMs (which could benefit, to some extent, from embedding CoT reasoning as part of the prompt), *o1* performs CoT reasoning internally. According to various benchmarks conducted by OpenAI, *o1* excels in areas such as programming, solving mathematical problems, and tackling complex puzzles, tasks that benefit from the more structured reasoning typically associated with CoT prompting [37].

Although the details of the functioning of *o1* are proprietary, it is known that *o1* has been trained using reinforcement learning, where it received feedback on the quality of its own reasoning behavior, thereby learning to undergo a thought process. The OpenAI website also mentions that a longer training period and a longer thinking process during output generation results in higher output quality. Thus, the performance of the model does not solely scale with the size of the base model, but also with the amount of computational power applied during training and inference. This new form of scaling could have major implications, as it may be easier to realize, for example, through faster computers or improvements in algorithms.

On its website, OpenAI provides several benchmark results and shows that *o1* particularly excels in mathematical tasks. The website states the following: "*We evaluated math performance on AIME, an exam designed to challenge the brightest high school math students in America. On the 2024 AIME exams, GPT-4o only solved on average 12% (1.8/15) of problems. o1 averaged 74% (11.1/15) with a single sample per problem, 83% (12.5/15) with consensus among 64 samples, and 93% (13.9/15) when re-ranking 1000 samples with a learned scoring function. A score of 13.9 places it among the top 500 students nationally and above the cutoff for the USA Mathematical Olympiad.*" [37]

The impressive performance of LLMs on mathematics benchmarks warrants further independent validation, particularly since some existing benchmarks have been criticized for their potential inadequacy [39,40]. For example, certain benchmarks rely on tasks such as multiple-choice questions, where an LLM could potentially 'cheat' by recalling answers from its training data rather than genuinely reasoning through the problems. To address these concerns, we conducted an independent evaluation of the *o1-preview* model by having it complete the national mathematics secondary school leaving examination in the Netherlands. In this paper, we report the results of the evaluation and compare the *o1-preview* model performance to that of the state-of-the-art *GPT-4o* model, which lacks advanced reasoning capabilities.

## 2. Method

### 2.1. Completing the Mathematics Exam

We used the exam 'VWO Mathematics B 2023' [41], the official national exam in abstract mathematics used in Dutch high schools. VWO stands for "voorbereidend wetenschappelijk onderwijs" ("preparatory scientific education"), which is the highest level of secondary education in the Netherlands. In a previous publication [42], we applied the same method to English comprehension exams and found that *GPT-4* scored between 8.1 and 8.3 on a scale of 1 to 10, placing it among the top 10% of human students who took the exam.

In this study, we used the following LLMs: *o1-preview-2024-09-12* (a model with System 2-like reasoning capabilities) and *gpt-4o-2024-05-13* (a model without these reasoning capabilities). OpenAI uses a parameter called 'temperature' that determines the degree of randomness in the output. A value close to 0 means that the token generator operates in an almost deterministic manner by producing the most likely token. A higher temperature value flattens the probability distribution of the tokens, which increases the likelihood of selecting other tokens. The temperature value for *GPT-4o* can be set between 0 (nearly deterministic output) and 2 (highly random, unreadable output), with a default value of 1. However, for *o1-preview*, the temperature value cannot be adjusted and is fixed at 1 by OpenAI, as this model is still in its beta development phase. To allow for a fairer comparison, we also ran *GPT-4o* with the default setting of a temperature of 1.

The selected final exam consisted of 19 questions and was first input as a textual prompt. Note that, unlike the human candidates, the LLMs did not have access to the images, as *o1-preview* is not multi-modal. The exam, and thus the prompts we provided to the LLMs, were in Dutch. Example prompts are provided below for Questions 11, 14, and 16:

*The function f is given by f(x) = abs(sin(x) + 1/2*sqrt(3)).* ~~In the figure, the graph of f is represented as a solid line. In the figure, the peaks A and B of the graph of f are indicated.~~ *A and B are the peaks corresponding to the first two maxima of f to the right of the y-axis. There exists a sinusoid given by*

*g(x) = a + bsin(x), for which two consecutive peaks coincide with points A and B. ~~The graph of g is shown as a dashed line in the figure.~~ Calculate the exact values of a and b*

*The function f is given by: f(x) = ln(x). The function g is given by: g(x) = 1 + e^2 * (1 - ln(x)). ~~In Figure 2, the graphs of f and g are shown again. For a certain value of q, the line with the equation y = q is also shown. This line~~ **The line** y = q intersects the graph of g at point A and the graph of f at point B, where point A lies to the left of point B. It is given that AB = 3. Calculate the corresponding value of q. Give your final answer to one decimal place.*

*Given is rectangle OABC with O(0,0), A(8,0), and C(0,4). Points F and E are the midpoints of OA and BC, respectively. Point P(0, p) lies on the negative y-axis. Point D is the intersection of the extension of line segment PF and line segment AC. The line through E and F is the bisector of angle PED. M(4, 2) is the intersection point of AC and EF. Circle c has center M and passes through D. Depending on the position of point P (and thus the value of p), the circle becomes larger or smaller. There is exactly one value of p for which circle c is tangent to both OA and BC. ~~In Figure 2, this situation is depicted.~~ Calculate this exact value of p.*

In the text above, the strikethrough indicates the text that was omitted (because the LLMs did not have access to the figures), and the boldface text reflects the adjustments made compared to the original exam.

Each of the 19 prompts was presented individually to the two LLMs using MathWorks MATLAB (version R2024a) via an API. The output was then submitted to a colleague (second author), who was asked to evaluate it according to the official answer key, which was available online [43].

**2.2. Repeating the Mathematics Exam**
Upon repeated prompting of several questions, we observed that the output of *o1-preview* and *GPT-4o* exhibited some variability, with the answers sometimes being incorrect where they had previously been correct. To investigate this variability more thoroughly, we had both models retake the exam using the same prompts as before.

Furthermore, for three selected questions, we conducted 20 repetitions to evaluate the degree of variability in the responses and determine if the self-consistency "consensus" method is effective [37,44,45].

**2.3. Completing a New Mathematics Exam**
The reported knowledge cutoff for the trained *o1-preview* and *GPT-4o* models was October 2023 [46], which introduced a risk of contamination. This means that our evaluation could be compromised if the answer key was seen during the model training process. Therefore, we repeated the above procedure, but this time used the Mathematics B exam 2024. Since this exam was only published online in May 2024, it is impossible for such contamination to have occurred. Additionally, we had this new exam completed by *o1-mini-2024-09-12*, an additional model with available API access, which, according to OpenAI, performs well and quickly on mathematics tasks.

Additionally, for one specific question, which *o1-preview* answered incorrectly, we repeated the prompting process 250 times. Again, the goal of this re-prompting was to assess whether the previously mentioned self-consistency method could effectively lead to the correct answer.

## 3. Results
**3.1. Completing the Mathematics Exam**
The results for *o1-preview* and *GPT-4o* are shown in Table 2. *o1-preview* achieved a perfect score of 76 out of 76. *GPT-4o*, the state-of-the-art model without reasoning features, achieved a score of 66 out of 76. For comparison, the 16,414 students in the Netherlands who took the same exam scored an average of 40.63 out of 76 [47]. The performances of *o1-preview* and *GPT-4o* compared to all the students who completed this exam are shown in Figure 1 (i.e., the first attempt).

From Table 2, it can also be observed that *o1-preview* used substantially more tokens (a total of 53,971) than *GPT-4o* (a total of 16,071). The input tokens were not counted here. Additionally, *GPT-4o* was faster (160.8 s) than *o1-preview* (617.2 s) at completing the 19 questions on the exam. The visible output of *o1-preview* was slightly more concise than that of *GPT-4o* (13,139 'non-reasoning' tokens compared to 16,071 tokens). The token usage of *o1-preview* was largely attributable to reasoning processes that were not visible in the output.

Table 2. *Results for all 19 questions of the Mathematics B final exam.*

| Question Number | Maximum Points | Mean Points Students | o1-Preview Points | GPT-4o Points | o1-Preview Number of Completion Tokens (of Which Reasoning) | GPT-4o Number of Completion Tokens | o1-Preview Completion Time (s) | GPT-4o Completion Time (s) |
|---|---|---|---|---|---|---|---|---|
| 1 | 3 | 2.36 | 3 | 3 | 811 (55.2%) | 745 | 11.5 | 6.9 |
| 2 | 5 | 4.15 | 5 | 5 | 1273 (50.3%) | 702 | 14.3 | 7.4 |
| 3 | 4 | 2.24 | 4 | 2 | 5733 (88.2%) | 1066 | 47.5 | 12.2 |
| 4 | 5 | 1.75 | 5 | 3 | 2373 (64.7%) | 881 | 56.2 | 10.8 |
| 5 | 3 | 2.63 | 3 | 3 | 993 (58.0%) | 561 | 11.7 | 7.0 |
| 6 | 4 | 1.48 | 4 | 4 | 2185 (70.3%) | 782 | 26.7 | 8.7 |
| 7 | 3 | 2.22 | 3 | 3 | 1219 (47.3%) | 686 | 16.2 | 6.2 |
| 8 | 3 | 2.08 | 3 | 3 | 2421 (66.1%) | 1181 | 26.2 | 9.9 |
| 9 | 3 | 1.64 | 3 | 3 | 2181 (55.8%) | 934 | 25.2 | 8.2 |
| 10 | 4 | 1.85 | 4 | 4 | 1999 (67.2%) | 837 | 22.9 | 7.9 |
| 11 | 3 | 1.20 | 3 | 3 | 5396 (84.2%) | 1062 | 68.5 | 10.3 |
| 12 | 5 | 3.56 | 5 | 4 | 2744 (70.0%) | 1168 | 30.8 | 13.1 |
| 13 | 6 | 3.51 | 6 | 6 | 1604 (51.9%) | 1078 | 19.3 | 10.3 |
| 14 | 4 | 2.13 | 4 | 2 | 6367 (92.5%) | 632 | 70.6 | 6.0 |
| 15 | 5 | 2.05 | 5 | 5 | 3908 (67.1%) | 950 | 40.4 | 8.9 |
| 16 | 6 | 0.94 | 6 | 3 | 8338 (92.1%) | 852 | 79.9 | 9.7 |
| 17 | 3 | 1.92 | 3 | 3 | 667 (57.6%) | 437 | 7.1 | 4.1 |
| 18 | 3 | 1.23 | 3 | 3 | 1947 (69.0%) | 702 | 21.9 | 5.8 |
| 19 | 4 | 1.69 | 4 | 4 | 1812 (60.0%) | 815 | 20.5 | 7.2 |
| **Total** | **76** | **40.63** | **76** | **66** | **53,971 (75.7%)** | **16,071** | **617.2** | **160.8** |

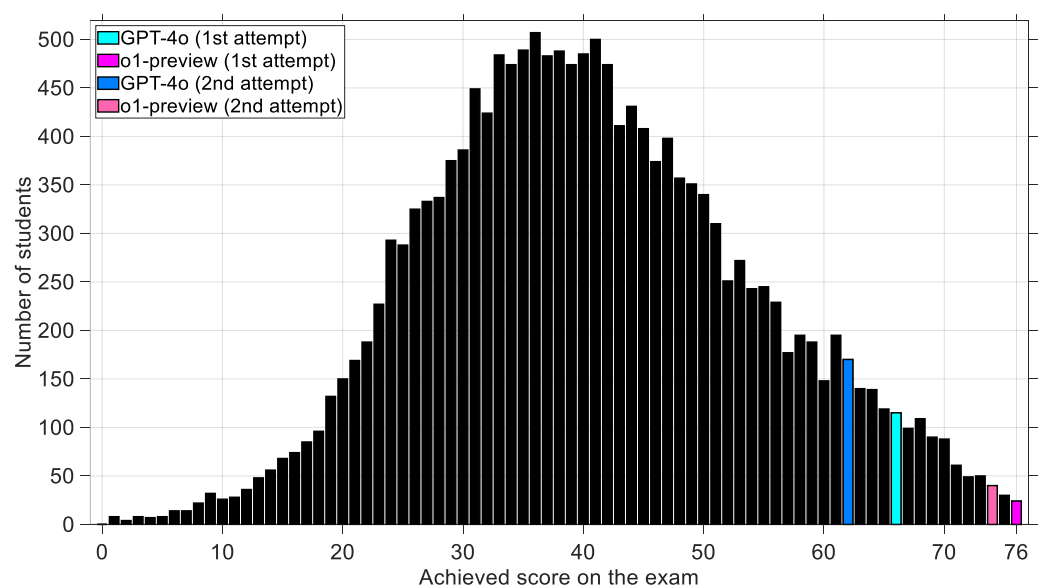

*Figure 1.* The distribution of the scores achieved by the students on the Mathematics B exam 2023 in the Netherlands (*n* = 16,414). The performances of *GPT-4o* and *o1-preview* are displayed as colored bars. Source of student data: Cito [48].

### 3.2. Repeating the Mathematics Exam
Despite the perfect score of *o1-preview*, some caveats remain. One factor to consider is that the perfect score was not consistently reproducible, possibly due to the non-zero temperature setting enforced by OpenAI, which introduced variability.

When we repeated the exam in the same manner, *o1-preview* and *GPT-4o* scored 74 out of 76 and 62 out of 76 points, respectively (see Figure 1, second attempt). Specifically, regarding *o1-preview*, the model now made a mistake on Question 16, scoring 4 out of 6 points, which was also the question the students found most difficult.

When repeating Question 16 twenty more times, *o1-preview* provided 12 correct answers and 8 incorrect answers. Other questions that exhibited variability included Question 11, which yielded 5 mistakes out of 20, and Question 3, which yielded 6 mistakes out of 20. These findings, where most of the responses were correct but some mistakes occurred, suggest that *o1-preview* could benefit from the self-consistency method [37,44,45], in which multiple answers are generated and the most common or "consensus" answer is selected. This strategy has also been acknowledged by OpenAI, as indicated in the introduction of the current paper.

### 3.3. Completing a New Mathematics Exam

The OpenAI website claims that "*The OpenAI o1-preview and o1-mini models share the same knowledge cut-off as our GPT-4o models, October 2023*" [46].

Since the above exam was taken on 11 May 2023, and the answer model appeared online not long after, this could pose an issue related to the phenomenon of contamination. In other words, it is possible that OpenAI scraped the internet and encountered the answers during the training or fine-tuning of the *GPT-4o* and *o1-preview* models. To mitigate this risk, a new evaluation was conducted using a Mathematics B exam taken on 23 May 2024.

We also expanded the evaluation with *o1-mini*, a smaller LLM from OpenAI that is designed to operate faster and more cost efficiently. According to OpenAI, *o1-mini* is particularly good at mathematical tasks and other STEM tasks, as long as the task does not require extensive world knowledge. This would also apply to math exams. The model we used here was *o1-mini-2024-09-12*.

For this new exam, the same procedure was followed as for the exam from 2023. The Mathematics B exam 2024 consisted of 18 questions, for which a total of 76 points could be earned.

The grading of the 2024 exam resulted in 71 points for *o1-preview*, 60 points for *GPT-4o*, and 72 for *o1-mini*. This means that the respective models scored in the 97.8th percentile (i.e., 12,842 out of 13,134 candidates achieved this score or lower), the 89.1st percentile (11,702 out of 13,134), and the 98.3rd percentile (12,910 out of 13,134), respectively [49].

The total time used was 1116.3 s for *o1-preview*, 163.4 s for *GPT-4o*, and 384.0 s for *o1-mini*. This confirms the above claim that o1-mini produces output faster. The number of completion tokens was 73,448 for *o1-preview*, 13,231 for *GPT-4o*, and 52,571 for *o1-mini*, with 81.6% of these being reasoning tokens for *o1-preview*, and 77.5% being reasoning tokens for *o1-mini*.

It should be noted that for *o1-preview* and *o1-mini*, we deducted 1 point for a rounding error (the produced answer for one question was 0.683 or 0.681, instead of 0.682), and for another question (Question 9), we found that, with repeated prompting, *o1-preview* gave the correct answer 89% of the time. This error could, thus, have been avoided by using the self-consistency approach. In short, even though *o1-preview* and *GPT-4o* scored slightly lower on the 2024 exam compared to the 2023 exam, this seems more like a coincidence, and there is no indication that OpenAI's knowledge cutoff explains this.

Figure 2 shows, for 249 repetitions of the answer to Question 9, how long *o1-preview* spent calculating and how this correlated with the number of reasoning tokens used. For this question, it turned out that incorrect answers were also often associated with longer-than-average reasoning times (on average, 151 s for incorrect answers versus 32.7 s for correct answers).

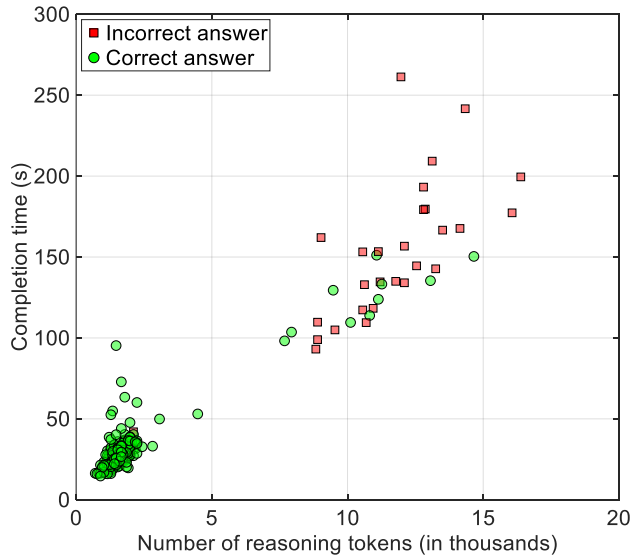

*Figure 2.* The completion time versus the number of reasoning tokens used by *o1-preview* for Question 9 on the Mathematics B exam 2024. The prompting was repeated 250 times, and one correct outlier response (completion time of 828 s; 12,160 reasoning tokens used) was removed. It can be observed that most of the answers were correct (221 of the 249 displayed markers are green), which implies that the self-consistency approach could be used to identify the correct answer.

## 4. Discussion

### 4.1. Main Results

This study shows that a new model from OpenAI, called *o1-preview*, which exhibits System 2-like reasoning abilities, achieved a near-perfect score on a Mathematics B final exam for high school students. The scores achieved by *o1-preview* were generally consistent with the results of various benchmark tests reported by OpenAI [13], and these have now been independently validated.

The scores for *o1-preview* of 76 out of 76 (Exam 2023, Attempt 1), 74 out of 76 (Exam 2023, Attempt 2), and 71 out of 76 (Exam 2024) translate to grades of 10.0, 9.9, and 9.7, respectively, something that only a few candidates in the Netherlands achieved. For the 2023 exam, in the whole Netherlands, there were 24 students with 0 mistakes (a score of 76 out of 76), 30 students with 1 mistake (a score of 75 out of 76), and 40 students with 2 mistakes (a score of 74 out of 76).

*GPT-4o*, the model without reasoning capabilities, also performed well, with scores of 66, 62, and 60 out of 76. In total, 96.1%, 93.0%, and 89.1% of the students in the Netherlands achieved these scores or lower, respectively. Finally, in the evaluation we conducted of *o1-mini*, a smaller model with System 2 reasoning capabilities, we found that it performed well, though not perfectly, achieving the 98.3rd percentile on the 2024 exam.

The obtained performance of the LLMs is quite conservative because, firstly, we presented the prompts in Dutch and not in English. Given that most of the training data are in English, the use of Dutch likely did not have a positive effect. See, for example, Yan et al. [50], for a study where this plays a role in the case of translation. However, the current study was in the field of mathematics, which is generally not very language intensive, so the effect of language was likely minimal. On the other hand, some of the questions had fairly extensive textual introductions related to some world knowledge.

Additionally, these high scores were obtained for a text-only input format, meaning that the models did not have access to the figures on the exam. For some questions, this meant that the models had to 'visualize' the spatial relationship between lines and points. The Mathematics B exam in the Netherlands is taken by VWO students, the highest level of secondary education in the country, and approximately 45% of VWO students took this exam [51]. Mathematics B is typically chosen by students who are more oriented toward STEM subjects. This makes the high performance of *o1-preview* all the more impressive. It should also be noted that, according to OpenAI, *o1-preview* is not the best model. There also exists an *o1* model (without the "*preview*" suffix) that offers vision capabilities and achieves better scores on benchmarks, but which is not yet accessible to the general public [37].

Another reason that the current results should be considered conservative is that we used only a single prompt per question. As we showed for the questions where *o1-preview* made mistakes, these mistakes could have been prevented by prompting multiple times and selecting the most common answer. Moreover, Figure 2 shows that the time or tokens used are indicative of whether a question was answered correctly or incorrectly by *o1-preview*. It should be noted, though, that while this self-consistency "consensus" method is effective, it is not particularly practical at the moment. The reason for this is that LLMs use a substantial amount of processing time (a median of 20 to 30 s per question for *o1-preview*, and about 10 s per question for *GPT-4o* and *o1-mini*). Given this limited inference speed, prompting multiple times would not be desirable for many applications. However, this may change in the future with the introduction of faster GPUs or better inference techniques. Whether the information about reasoning time, as shown in Figure 2, is useful for converging on a correct answer deserves further research and validation. It will not apply to every question. For example, it could be the case that for more difficult questions, a short reasoning time rather than a long reasoning time is actually indicative of an incorrect answer.

### 4.2. Outlook and Conclusions

The fact that *o1-preview* can perfectly complete a final exam implies that people now have access, via an API or web interface, to a tool that can solve problems at an extraordinarily high level and at superhuman speed. *o1-preview* completed the entire 2023 exam in 10.3 min compared to 2.7 min for *GPT-4o*, while the human candidates were given 180 min. The OpenAI website suggests that the performance of this new System 2-like method of prompting scales well to higher levels of accuracy. This raises important questions regarding the implications of this technology. In particular, we wonder how science will adapt, especially considering that problem solving by AI will become faster and better in the future. A similar question arises for education, where the role of a human teacher is called into question now that a generic problem solver and step-by-step explainer is available via a laptop or mobile phone.

The high performance of the *o1-preview* model also brings safety concerns. If the model is capable of solving difficult problems, as demonstrated in this work, it could potentially also be used to solve malicious tasks. For example, an advanced model might identify vulnerabilities in a computer system and exploit them (e.g., hacking), or generate harmful software (viruses) designed to cause maximum damage (see OpenAI o1 System Card for a detailed risk assessment [52]). On the other hand, LLMs with System 2-like reasoning could revolutionize problem-solving in domains such as medicine and engineering, or could offer advanced personalized education or decision-making support. In a different context, an AI-powered robot with System 2-like reasoning could optimize real-time decision making in industrial environments, such as factory floors or warehouses. Such a system could manage goods, reconfigure production lines to meet fluctuating demand, and respond efficiently to equipment failures or supply chain disruptions.

We are considering using language reasoning models in mechanical engineering education and workshops, for example, in subjects like strength theory and statics. An LLM-based software tool could potentially be used to assess which reasoning errors a student makes and tailor explanations accordingly. At the same time, it should be noted that the current evaluation was conducted on high school exams. Although these were challenging math problems, with very few students in the Netherlands solving all the questions flawlessly, the level is still limited. Some evaluations that appeared on preprint servers after we had submitted our paper offer further insights into the possibilities and limitations of *o1-preview* and *o1-mini* [53,54]. In particular, Fan et al. [54] showed that *o1-mini* (5-shot CoT prompting), although it outperformed the other LLMs tested, achieved a score of only 62.3% on HARDMath, a benchmark dataset of advanced applied mathematics problems.

In conclusion, the current work has demonstrated that a new generation of LLMs capable of reasoning (System 2 processes) has led to a substantial improvement in their performance on mathematical problems compared to the previous generation of LLMs. This new technology raises questions about the extent to which AI can mimic or even surpass human cognitive processes. It also prompts questions about the future of scientific research, education, safety, and potential applications.

## Data Availability Statement

The MATLAB script that was used to call the API and the files with all the results and evaluation of the output are available via the following link: https://doi.org/10.4121/2e663686-f656-4ff2-bb21-567ba4d4f03e.v2.

## Conflicts of Interest

The authors declare no conflicts of interest.